\documentclass[a4paper,11pt]{article}
\usepackage[utf8]{inputenc}
\usepackage[T1,T2A]{fontenc}
\usepackage{amsfonts}
\usepackage{adjustbox}
 \usepackage{amsmath}
\usepackage{amssymb}
\usepackage{graphicx}
\usepackage{wrapfig}
\usepackage{hyperref}
\usepackage{verbatim}
\usepackage[dvipsnames]{xcolor}
\usepackage{subcaption}
\usepackage{graphicx}
\usepackage{cite}

\begin{document}

\begin{center}
{\LARGE {\bf Warm fermionic dark matter from freeze-in at stronger coupling}}
\end{center}

\vspace{1cm}
\begin{center}
{\bf   Duarte Feiteira$^1$, Vinícius Oliveira$^{2,3}$}
\end{center}

\begin{center}
  \vspace*{0.25cm}
  $^1$ Department of Physics and Helsinki Institute of Physics, Gustaf Hallstromin katu 2a, FI-00014 Helsinki, Finland\\
  $^2$Departamento de Física da Universidade de Aveiro, Campus de Santiago, 3810-183 Aveiro, Portugal \\
 $^3$Laboratório de Instrumentação e Física Experimental de Partículas (LIP), Universidade do Minho, 4710-057 Braga, Portugal 
\end{center}

\vspace{2.5cm}

\begin{center} {\bf Abstract} \end{center}
\noindent
We study warm fermionic dark matter (DM) in the framework of freeze-in at stronger coupling, in the minimal Higgs portal scenario.
%with a singlet Majorana fermion. 
The reheating temperature is taken to be low, so that DM production from the Standard Model thermal bath is Boltzmann-suppressed and the DM stays out of equilibrium even for a sizeable coupling. This opens the possibility of observable signatures, in particular invisible Higgs decays. We compute the DM relic abundance including both the pre- and post-reheating contributions.  We find that the fermionic DM production reaction is strongly velocity suppressed, requiring larger reheating temperatures than those obtained for scalar DM in order to reproduce the correct relic abundance. The resulting DM momentum distribution is strongly non-thermal and its shape is not captured by the common $\alpha\beta\gamma$-parametrization. We find that the Lyman-$\alpha$ constraint excludes DM masses below about $100 -180\,\mathrm{keV}$, depending on the reheating history.

\vspace{1cm}

\newpage 
\tableofcontents

%%%%%%%%%%%%%%%%%%
\section{Introduction}
%%%%%%%%%%%%%%%%%%

The nature of dark matter (DM) remains one of the central open questions in physics \cite{Planck:2018vyg}. In the standard weakly interacting massive particle (WIMP) picture, DM reaches thermal equilibrium with the Standard Model (SM) plasma and its abundance is set by freeze-out \cite{Kolb:1990vq,Gondolo:1990dk}, which requires a sizeable coupling to the visible sector. The absence of experimental confirmation has motivated non-thermal alternatives. The leading one is the freeze-in mechanism \cite{Hall:2009bx}, in which DM is slowly produced from the thermal bath through a feeble coupling and never thermalizes. The price of this mechanism is that the same small coupling makes the scenario very hard to be probed experimentally.

This conclusion can be evaded in a low-reheating temperature scenario \cite{Cosme:2024ndc}. In this case, the production rate is Boltzmann-suppressed and DM never thermalizes, even for a sizeable coupling to the SM \cite{Lebedev:2024mbj}. A low $T_R$ is also welcome from the top-down perspective: it suppresses the relics produced gravitationally during reheating \cite{Lebedev:2022ljz,Koutroulis:2023fgp}, which otherwise may \textit{contaminate} freeze-in predictions \cite{Lebedev:2022cic}, so that a negligible initial DM abundance is a consistent assumption. Motivated by this, we focus on reheating temperatures below $800\,\mathrm{MeV}$.

In the Higgs portal framework, the SM Higgs boson mediates between the visible and dark sectors. A significant Higgs–DM coupling then leads to invisible Higgs decays, which both constrain the model and offer a way to test it at the LHC and at future collider experiments. This has been worked out for DM masses around and above the MeV scale \cite{Lebedev:2024mbj,Cosme:2023xpa,Koivunen:2024vhr,Arcadi:2024wwg}, where DM is \textit{cold}. Sub-MeV DM, however, is produced with a large comoving momentum and is therefore warm. Its free-streaming may erase small-scale structure, and the resulting Lyman-$\alpha$ bound becomes the dominant constraint. A central feature of low-temperature freeze-in is that the production is cutoff at the finite temperature $T_R$, which leaves the DM momentum distribution highly non-thermal and makes such structure-formation bounds qualitatively different from the standard freeze-in case. The scalar DM Higgs portal in this warm regime was studied in Ref.~\cite{Feiteira:2026qme}. In this work, we extend that analysis to fermionic DM.

%%%%%%%%%%%%%%%%%%%%%%%%%%%%
\section{Theoretical setup}
\label{Sect:theory}
%%%%%%%%%%%%%%%%%%%%%%%%%%%%
We consider a minimal Higgs portal scenario with a singlet fermionic dark matter candidate (see Refs.~\cite{Patt:2006fw,Lebedev:2021xey} for a review). Both CP-even and CP-odd couplings between the DM and the Higgs field are taken into account separately. The interactions between the SM Higgs $\mathrm{SU}(2)_W$-doublet $\mathcal{H}$ and DM are given by
\begin{align}
    \mathcal{L}_{h \chi} &= \frac{1}{\Lambda} \mathcal{H} \mathcal{H}^\dagger \chi \bar \chi \,, \\ 
    \mathcal{L}_{h \chi}^{\gamma_5} &= \frac{1}{\Lambda_5} \mathcal{H} \mathcal{H}^\dagger  \chi  i \gamma_5 \bar \chi \,,
\end{align}
where $\chi$ is a \textit{Majorana} fermion with mass $m_\chi$, assumed to be stable and thus a viable DM candidate. Since these interactions are non-renormalizable, they are described within an effective field theory framework in terms of the scales $\Lambda$ and $\Lambda_5$ \cite{Kim:2006af,Lopez-Honorez:2012tov}.

We focus on energy scales below $m_h/2$, such that the physical Higgs boson $h$ (with $m_h \simeq 125~\mathrm{GeV}$) can be integrated out. Using the SM Yukawa interaction
\begin{equation}
    \mathcal{L}_\mathrm{SM} = \frac{m_f}{v} \, h f\bar f \,,
\end{equation}
with $v \simeq 246~\mathrm{GeV}$, one obtains the effective four-fermion operators
\begin{align}
    \mathcal{L}^{\chi}_\mathrm{eff} &= \frac{m_f}{\Lambda m_h^2} f \bar f \,  \chi \bar \chi \,, \\ 
    \mathcal{L}^{\chi \, \gamma_5}_\mathrm{eff} &= \frac{m_f}{\Lambda_5 m_h^2} f \bar f \,  \chi  i \gamma_5 \bar \chi \,.
\end{align}
These operators mediate DM production from the SM thermal bath at low temperatures.

\begin{figure}
    \centering
    \includegraphics[width=0.35\linewidth]{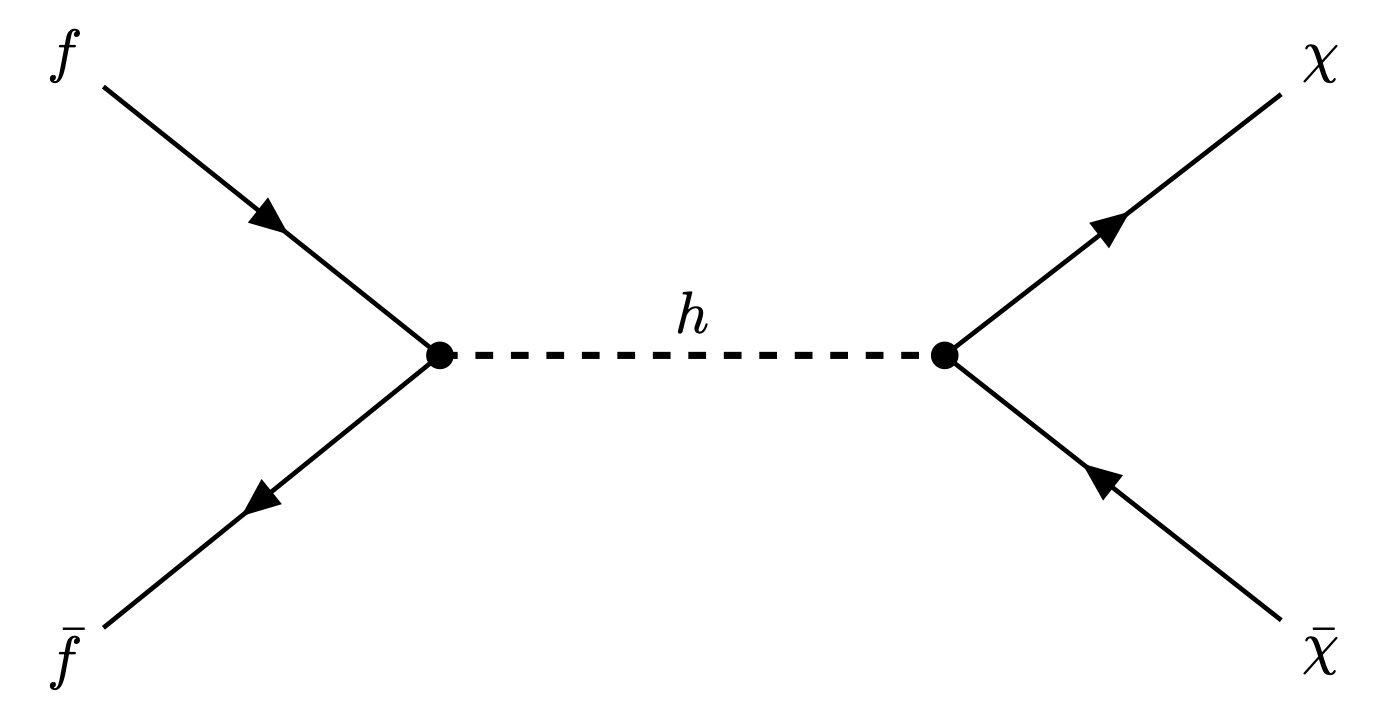}
    \caption{Main contributions to fermionic DM production at low temperatures.}
    \label{fig:DM_production}
\end{figure}

In the low reheating temperature framework, DM is produced via annihilation of SM particles in the thermal bath, as described in \autoref{fig:DM_production}. The production rate can be reliably computed perturbatively above the QCD confinement scale, $T \gtrsim 150~\mathrm{MeV}$. At lower temperatures, however, the dynamics become non-perturbative due to hadronization, and one must in principle account for hadronic initial states. However, as will be shown later, the fermionic DM scenario considered here typically favors reheating temperatures above the confinement scale in non-constrained region of the parameter space.

The spin-averaged cross sections for $f\bar f \to \chi \bar \chi$ for CP-even and CP-odd couplings are given by
\begin{align}
    \sigma(f \bar f \to \chi \bar \chi) &= \frac{N_c}{2}\,\frac{m_f^2}{4\pi \Lambda^2 m_h^4 s} \sqrt{s-4m_f^2}\,(s-4m_\chi^2)^{3/2} \,, \\ 
    \sigma(f \bar f \to \chi \bar \chi) &= \frac{N_c}{2}\,\frac{m_f^2}{4\pi \Lambda_5^2 m_h^4 } \sqrt{s-4m_f^2}\,(s-4m_\chi^2)^{1/2}\,,
\end{align}
respectively, where $s$ is the Mandelstam variable,  the factor $1/2$ accounts for identical particles in the final state, $N_c = 3$ for quarks and $N_c = 1$ for leptons are the colors  factor.

In the limit $m_\chi \ll m_f$, relevant for sub-MeV DM, both cross sections reduce to
\begin{equation}\label{eq:CS}
    \sigma(f \bar f \to \chi \bar \chi) = \frac{ N_c m_f^2}{8\pi \Lambda^2 m_h^4} s\,\sqrt{1-\frac{4m_f^2}{s}}\,.
\end{equation}
As we will demonstrate below, the explicit $s$-dependence of the cross section leads to a preference for reheating temperatures above the QCD phase transition scale.

In the low reheating temperature regime, sizeable Higgs--DM couplings can be accommodated without thermalizing the dark sector \cite{Lebedev:2024mbj}. This opens the possibility of probing the model via Higgs decays into invisible final states, which set important constraints on the effective scales $\Lambda$ and $\Lambda_5$. Current LHC searches constrain the invisible Higgs branching ratio to $\mathrm{BR}_\mathrm{inv} \lesssim 10\%$~\cite{ATLAS:2023tkt}, with a projected sensitivity of $\mathrm{BR}_\mathrm{inv} \lesssim 3\%$ at the HL-LHC~\cite{RivadeneiraBracho:2022sph}. The FCC benchmark goal has been considered in earlier studies as $\mathrm{BR}_\mathrm{inv} = 0.3\%$ \cite{FCC}. However, recent studies indicate that a better sensitivity can be achieved, with larger integrated luminosities, leading to a benchmark of $\mathrm{BR}_\mathrm{inv} = 0.05\%$ \cite{FCC2}.  
The Higgs decay widths into CP-even and CP-odd $\chi$ are given by
\begin{align}
    &\Gamma(h\to \bar{\chi} \chi) = \frac{m_h}{4\pi} \frac{v^2}{\Lambda^2} \left( 1 - \frac{4m_\chi^2}{m_h^2} \right)^{3/2}\,, \\
    &\Gamma(h\to \bar{\chi} \chi) = \frac{m_h}{4\pi} \frac{v^2}{\Lambda_5^2}
    \left( 1 - \frac{4m_\chi^2}{m_h^2} \right)^{1/2}\,,
\end{align}
respectively, where the symmetry factor $1/2$ has been included. In the limit $m_\chi \ll m_h$, both expressions reduce to
\begin{equation}\label{eq:Decay}
    \Gamma(h\to \bar{\chi} \chi) \approx \frac{m_h}{4\pi} \frac{v^2}{\Lambda^2}\,.
\end{equation}

In the sub-MeV DM limit, the cross sections and decay widths for CP-even and CP-odd interactions coincide. Therefore, constraints derived in one case can be directly applied to the other. This behaviour was also observed in Ref.~\cite{Lebedev:2024mbj}. In what follows, we thus focus on Eqs.~\ref{eq:CS} and \ref{eq:Decay}. The results obtained will apply to both scenarios.

%%%%%%%%%%%%%%%%%%%%%%%%%%%%
\section{Temperature evolution in the Early Universe}
%%%%%%%%%%%%%%%%%%%%%%%%%%%%

In this work, we consider the thermal history of the Universe developed in Ref.~\cite{Cosme:2024ndc}. In this section we limit ourselves to summarizing how the temperature evolves in such scenario. The thermal history of the SM bath prior to reheating depends on how the visible sector is populated after inflation. A straightforward possibility is that the decay of some primordial field produced the visible sector, which can be the inflaton $\phi$  \cite{Kolb:1990vq} or some intermediate component (e.g. a right-handed neutrino $\nu_R$, through $\phi \to \nu_R \nu_R\to {\rm SM}$), whose energy density redshifts as $\rho \propto a^{-n}$, while the Hubble rate scales as $H \propto a^{-m}$. As long as the bath sourced by its decay does not dominate the energy budget, one has $\rho_{\rm SM}\propto a^{-(n-m)}$, and, assuming instantaneous thermalization,
\begin{equation}
T_{\rm SM}(a) \propto a^{-\frac{n-m}{4}} .
\end{equation}
The reheating temperature $T_R$ is the SM temperature at the onset of radiation domination. The maximum temperature $T_{\max}$ reached by the SM bath before the reheating  may be larger than $T_R$\cite{Kolb:1990vq,Giudice:2000ex,Garcia:2020eof}.

The qualitative pre-reheating temperature behaviour depends if $n-m$ has sign or is zero. For direct inflaton decay $\phi \to \mathrm{SM}$, $m=n/2$ and $\rho_\mathrm{SM} \propto a^{-m}$, with $m$ being $3/2$ or $2$ for the non-relativistic and relativistic inflaton, respectively, and $ T_{\max} > T_R$ before the reheating. If $\phi$ instead feeds the SM bath through an intermediate component, for example via the decay chain $\phi \to \nu_R \nu_R\to {\rm SM}$, then $n-m=0$, and the SM temperature stays constant before reheating  $T_{\max}\simeq T_R$, realizing a \emph{flat-temperature profile}. A longer decay chain can give $n-m<0$, so that $T_{\rm SM}$ grows with time before the reheating, realizing the \emph{instant reheating profile}.

These histories affect both the DM relic abundance and its phase-space distribution. We analyze the two representative cases: the instant-like profile, dominated by production at reheating, and the flat-temperature profile, which receives a pre-reheating contribution.

%%%%%%%%%%%%%%%%%%%%%%%%%%%%
\section{Freeze-in at stronger coupling for instant reheating temperature profile}
%%%%%%%%%%%%%%%%%%%%%%%%%%%%

We first adopt the instant reheating approximation, in which the SM temperature rises abruptly from $T \simeq 0$ to $T_R$ and DM production begins at this point. The number density $n_\chi$ then obeys the Boltzmann equation
\begin{equation}
\dot{n}_\chi + 3 H n_\chi = 2 \, \Gamma_{f\bar f\to \chi \bar \chi} - 2 \, \Gamma_{\chi \bar \chi \to f\bar{f}}\,,
\end{equation}
where $H$ is the Hubble rate, the factor of $2$ signifies production or annihilation of two DM quanta per reaction, and $\Gamma_{f\bar f\to \chi \bar \chi}$, $\Gamma_{\chi \bar \chi \to f\bar{f}}$ are the production and annihilation rates per unit volume. Since DM is entirely produced after reheating, we set $n_\chi(T_R)=0$.

In the radiation-dominated era, the Hubble rate and entropy density read
\begin{equation}
    H = \sqrt{\frac{\pi^2 g_e(T)}{90}}\frac{T^2}{M_\mathrm{Pl}}\,, 
    \qquad
    s_{\mathrm{SM}} = \frac{2\pi^2}{45} g_s(T)T^3\,,
\end{equation}
where $g_e$ and $g_s$ count the SM degrees of freedom contributing to the energy and entropy densities, respectively, and $M_\mathrm{Pl} = 2.43 \times 10^{18}$ GeV is the reduced Planck mass. The production rate per unit volume is \cite{Gondolo:1990dk}
\begin{align}
\Gamma_{f\bar f\to \chi \bar \chi}
&= \langle \sigma v_r \rangle n_f^{\text{(eq)}2} \\
&= 2^2 \times \frac{T}{32 \pi^4} \int_{\text{max}(4m_f^2, 4m_\chi^2)}^\infty d s \,
\sigma(s) (s-4m_f^2)\sqrt{s} K_1 \left( \frac{\sqrt{s}}{T} \right)\,,
\end{align}
where the factor $2^2$ accounts for the spin d.o.f. of the initial fermions, $\langle\sigma v_r\rangle$ is the thermally averaged cross section times the relative velocity $v_r$, $n_f^{\text{(eq)}}$ is the equilibrium number density of $f$, and  $K_1(x)$ is the modified Bessel function of the first kind.

Parametrizing the abundance as $Y_\chi = n_\chi / s_{\mathrm{SM}}$ and trading $T$ for $x = m_\chi/T$, the Boltzmann equation becomes
\begin{equation}
    \frac{dY_\chi}{dx} = 
    2 \sqrt{\frac{8\pi^2 M_\text{Pl}^2}{45}} 
    \frac{g_*^{1/2} m_\chi}{x^2}
    \sum_{f}
    \langle \sigma v_r \rangle
    Y_f^{\text{(eq)}2}
    \left[1 - \left(\frac{Y_\chi}{Y_\chi^\text{eq}} \right)^2 \right]\,,
\end{equation}
with $Y_f^{\text{(eq)}} = n_f^\mathrm{(eq)}/s_\mathrm{SM}$ carrying the spin factor $g_f=2$ of the initial fermions and
\begin{equation}
    g_*^{1/2} = \frac{g_s}{g_e^{1/2}}\left( 1 + T \frac{dg_s/dT}{3 g_s} \right)\,.
\end{equation}
The observed relic density corresponds to
\begin{equation}
    Y_\infty = 4.4 \times 10^{-10} \frac{\mathrm{GeV}}{m_\chi}\,.
\end{equation}

\subsection{Simple estimates of the DM relic abundance}

In the freeze-in regime the backreaction term $\Gamma_{\chi \bar \chi \to f\bar{f}}$ is negligible, since the DM abundance stays far below its equilibrium value. In this case, the Boltzmann equation can be kept in the form
\begin{equation}\label{eq:pure_Freeze-in_Y}
    \frac{dY_\chi}{dx} = 
    2 \sqrt{\frac{8\pi^2 M_\text{Pl}^2}{45}} 
    \frac{g_*^{1/2} m_\chi}{x^2}
    \langle \sigma v_r \rangle
    Y_f^{\text{(eq)}2}\,,
\end{equation}
which we now evaluate in the two regimes set by the relation between $T_R$ and the fermion mass $m_f$: $T_R \ll m_f$, and $T_R \gg m_f$.

% =========================================================
\subsubsection{\texorpdfstring{$T_R \ll m_f$}{T\_R << m\_f}}
The reaction rate for a single Dirac fermion $f$ can be written as
\begin{equation}\label{eq:rate_reaction_estimation}
\Gamma_{f\bar f\to \chi \bar \chi}(T)
=
2^2 \times \frac{T}{32\pi^4}
\int_{4m_f^2}^{\infty}
ds\,
\sigma(s)
(s-4m_f^2)
\sqrt{s}
K_1\!\left(\frac{\sqrt{s}}{T}\right)\,,
\end{equation}
where we assumed the regime $m_f > m_\chi$ which sets the lower limit of $s$ at $4m_f^2$, and the cross section is given by~\autoref{eq:CS}. To compute the integral, we may use the large argument expansion $K_1(\sqrt{s}/T) \simeq \sqrt{\frac{\pi}{2}} T^{1/2}s^{-1/4} e^{-\sqrt{s}/T}$. The integral is dominated by the threshold $s \simeq 4m_f^2$, and therefore we may approximate the ''slow'' function $s^{3/4} \to (4m_f^2)^{3/4}$, which leaves a non-trivial integral of the form
\begin{equation}
\int_{4m_f^2}^\infty ds \, s \, (s-4m_f^2)^{3/2} e^{-\sqrt{s}/T}
\simeq
96 \sqrt{\pi}\, m_f^{9/2} T^{5/2} e^{-2m_f/T}\,,
\end{equation}
so that\footnote{Compared to the scalar case of Ref.~\cite{Feiteira:2026qme}, the fermionic contribution depends more steeply on the center-of-mass energy, which enhances the production rate despite the Boltzmann suppression.}
\begin{equation} \label{eq:RR_Boltz_Suppr}
\Gamma_{f\bar f\to \chi \bar \chi}
\simeq
\frac{3 N_cm_f^6}{4 \pi^4 \Lambda^2 m_h^4}\,
T^4\, e^{-2 m_f/T}\,.
\end{equation}

The Boltzmann equation \eqref{eq:pure_Freeze-in_Y} can then be solved analytically using $g_* \simeq 10$ and setting the
boundary condition $Y_\chi (T_R) =0$. The result is
\begin{equation}
Y_\chi
\simeq
\frac{81  N_c  m_f^5 M_\mathrm{Pl}}{16 \pi^7 \Lambda^2 m_h^4}
\, e^{-2 m_f/T_R}\,.
\end{equation}
Other choices of $g_*$ can be accommodated using the scaling $Y_\chi \propto g_*^{-3/2}$. The $T_R$ dependence is also shared by the scalar Higgs portal DM model \cite{Lebedev:2024mbj}.

% =========================================================
\subsubsection{\texorpdfstring{$T_R \gg m_f$}{T\_R >> m\_f}}

To compute the reaction rate, given by \autoref{eq:rate_reaction_estimation}, in such regime,  we define the new variable $s\to x^2 \ T^2$. In this regime the production peaks at $T_R$ and the fermions are relativistic, therefore we may approximate $4m_f^2/T \to 0$, which leaves an integral of the form
\begin{equation}
    \int_0^\infty dx \ x^6  K_1(x) = 384 \,.
\end{equation}
This yields
\begin{equation}\label{eq:rate_relativistic}
\Gamma_{f\bar f\to \chi \bar \chi}
\simeq
\frac{12 N_c m_f^2 T^8}{\pi^5 \Lambda^2 m_h^4}\,.
\end{equation}
We see that the rate is not suppressed by $e^{-2 m_f/T}$, since the fermions are relativistic.  The resulting DM abundance is
\begin{equation}
Y_\chi
\simeq
\frac{54 N_c m_f^2 M_\mathrm{Pl} T_R^3}{\pi^8 \Lambda^2 m_h^4}\,,
\end{equation}
with $g_* = 10$.

Because of the explicit $s$-dependence of the cross section, channels with $T_R \ll m_f$ stay effective down to lower temperatures than those with $T_R \gg m_f$. Heavier fermions can therefore dominate the production despite the Boltzmann suppression of their number density.

%%%%%%%%%%%%%%%%%%%%%%%%%%%%
\section{Freeze-in at stronger coupling for flat temperature profile}
%%%%%%%%%%%%%%%%%%%%%%%%%%%%

We now turn to the flat temperature profile, in which the SM bath sits at a constant temperature, $T_R$, throughout the pre-reheating stage. Unlike the instantaneous case, this scenario receives an additional contribution from the pre-reheating epoch, denoted $\Delta Y_\chi$. Using $dt =da/(aH)$ we can write the Boltzmann equation in terms of the scale factor at fixed $T = T_R$,
\begin{equation}\label{eq:Fla_T_abundance}
\frac{d\Delta Y_\chi}{da}
=
\frac{1}{a H(a) s_{\rm SM}(T_R)}
\left[
2\,\Gamma_{f\bar f\to \chi \bar \chi}(T_R)
-
3 H(a) s_{\rm SM}(T_R)\, \Delta Y_\chi
\right]\,,
\end{equation}
where we neglected the backreaction term $\Gamma_{\chi \bar \chi \to f\bar{f}}$. The first term in the bracket describes the continuous production of DM and the second its dilution by expansion. We will adopt $H \propto a^{-2}$ during this period, corresponding to a radiation-like energy density. This choice is not essential, and similar results follow for other scaling laws.

The Boltzmann equation \eqref{eq:Fla_T_abundance} can be solved analytically by setting $\Delta Y_\chi (0) =0$ and integrating from $0$ to $a_R$. The result is
\begin{equation}\label{eq:Delta_Y}
    \Delta Y_\chi(a_R)
    =
    \frac{2\Gamma_{f\bar f\to \chi \bar \chi}(T_R)}
    {5 H(T_R) s_{\rm SM}(T_R)}\,,
\end{equation}
which is the total yield accumulated during the constant-temperature phase. The final abundance is the sum of the pre- and post-reheating contributions,
\begin{equation}
    Y_\chi^\mathrm{Flat-T} = Y_\chi + \Delta Y_\chi(a_R)\,,
\end{equation}
with $Y_\chi$ given in the previous section.

\subsection{Simple estimates of the DM relic abundance}

As before, we will estimate $\Delta Y_\chi$, given by \autoref{eq:Delta_Y},  in the two different regimes set by the relation between $T_R$ and the fermion mass $m_f$: $T_R \ll m_f$, and $T_R \gg m_f$. The reaction rates used here are the same as those computed in the previous section.

\subsubsection{\texorpdfstring{$T_R \ll m_f$}{T\_R << m\_f}}

In this regime, the reaction rate for a single Dirac fermion $f$ is given by \autoref{eq:RR_Boltz_Suppr}, yielding
\begin{equation}
\Delta Y_\chi(a_R)
\simeq
\frac{81 N_c m_f^6 M_\mathrm{Pl}}{40 \pi^7 \Lambda^2 m_h^4 T_R}
\, e^{-2 m_f/T_R}\,,
\end{equation}
with $g_s \simeq g_e \simeq 10$.

\subsubsection{\texorpdfstring{$T_R \gg m_f$}{T\_R >> m\_f}}

Here the production is not Boltzmann-suppressed. The reaction rate is given by \autoref{eq:rate_relativistic}, yielding
\begin{equation}
\Delta Y_\chi(a_R)
\simeq
\frac{162 N_c m_f^2 M_\mathrm{Pl} T_R^3}{5 \pi^8 \Lambda^2 m_h^4}\,,
\end{equation}
with $g_s \simeq g_e \simeq 10$.

The relative weight of the two epochs follows from their ratio. For $T_R \ll m_f$,
\begin{equation}
    \frac{Y_\chi}{\Delta Y_\chi(a_R)}
    =
    \frac{5}{2}
    \frac{T_R}{m_f}\,,
\end{equation}
so production is dominated by the pre-reheating epoch at low $T_R$. For $T_R \gg m_f$, instead,
\begin{equation}
    \frac{Y_\chi}{\Delta Y_\chi(a_R)}
    = 5/3\,,
\end{equation}
independently of $T_R$, with the post-reheating period taking over.

%%%%%%%%%%%%%%%%%%%%%%%%%%%%
\section{Phase space distribution}
%%%%%%%%%%%%%%%%%%%%%%%%%%%%

The DM phase-space distribution $f_\chi(p,T)$ evolves according to the Boltzmann equation in an expanding Universe \cite{Ballesteros:2020adh},
\begin{equation}
\frac{\partial f_\chi}{\partial t} - H p \frac{\partial f_\chi}{\partial p} = \mathcal{C}[f_\chi] \,,
\end{equation}
where $H$ is the Hubble rate and $\mathcal{C}[f_\chi]$ the collision term encoding particle production. In general $\mathcal{C}[f_\chi]$ is a functional of the DM distribution, as it accounts for both production and annihilation. In the freeze-in regime, however, $f_\chi$ lies far below equilibrium and the back-reaction $\chi\bar\chi\to f\bar f$ is negligible, so the collision term no longer depends on $f_\chi$. It reduces to an ordinary function of the DM momentum $p$ and the bath temperature $T$, which we denote $\mathcal{C}(p,T)$ and refer to as the collision term, set entirely by the equilibrium distributions of the bath. For $f\bar f \to \chi\bar\chi$ it reads \cite{DEramo:2020gpr}
\begin{equation}\label{eq:Collision}
    \mathcal{C}(p,T) =  \frac{1}{16\pi^3}\frac{T e^{-E/T}}{p} 
    \int_{4 m_f^2}^\infty ds \, \overline{|\mathcal{M}|^2} 
    \sqrt{1- \frac{4m_f^2}{s}} \, e^{-\frac{s}{4pT}}\,,
\end{equation}
where we neglected the DM mass since $m_\chi \ll m_f$, and the matrix element averaged over initial and final states is
\begin{equation}
    \overline{|\mathcal{M}|^2}_{f\bar f\to \chi\chi} = N_c\frac{4 m_f^2}{\Lambda^2 m_h^4} s(s-4m_f^2)\,.
\end{equation}

In terms of the comoving momentum $q \equiv p/T$, the distribution is given by \cite{Feiteira:2026qme}
\begin{equation}\label{eq:psd}
f_\chi(q) = \frac{1}{q} \int_{T_0}^{T_R} \frac{dT}{T^2} \frac{\mathcal{C}(q,T)}{H(T)}\,,
\end{equation}
with $T_0$ the present temperature. The key departure from standard freeze-in is that the integration is cut off at the finite reheating temperature $T_R$ instead of extending to high temperatures, which is what shapes the non-thermal distribution.

The shape of $f_\chi(q)$ is governed by the dominant part of the collision term,
\begin{equation}
    \mathcal{C}(p,T) \propto \frac{T e^{-p/T}}{p}
    \int_{4m_f^2}^\infty ds\, \sqrt{s}\, (s-4m_f^2)^{3/2}\, e^{-\frac{s}{4pT}}\,.
\end{equation}
Changing variables to $s = 4m_f^2 + 4pT x$ and evaluating the integral in the two regimes, one finds
\begin{align}
    &p \ll \frac{m_f^2}{T} \, : \qquad 
    \mathcal{C}(p,T) \propto m_f\, p^{3/2} T^{7/2} 
    \exp\left(-\frac{p}{T} - \frac{m_f^2}{pT}\right)\,, 
    \label{eq:Asymp_1} \\[4pt]
    &p \gg \frac{m_f^2}{T} \, : \qquad 
    \mathcal{C}(p,T) \propto p^{2} T^{4} 
    \exp\left(-\frac{p}{T} - \frac{m_f^2}{pT}\right)\,.
    \label{eq:Asymp_2}
\end{align}
The exponential is fixed by the collision kinematics: a soft DM particle of momentum $p$ is produced together with a hard one of momentum $\sim m_f^2/p$, both drawn from the Boltzmann tail of the bath. Compared to the scalar case \cite{Feiteira:2026qme}, the steeper powers of $p$ follow directly from the $s$-dependence of the fermionic amplitude, $\overline{|\mathcal{M}|^2}\propto s(s-4m_f^2)$.

We now apply this result to the two thermal histories.

% =========================================================
\subsection{Instant reheating temperature profile}

\begin{figure}
    \centering
    \includegraphics[width=0.5\linewidth]{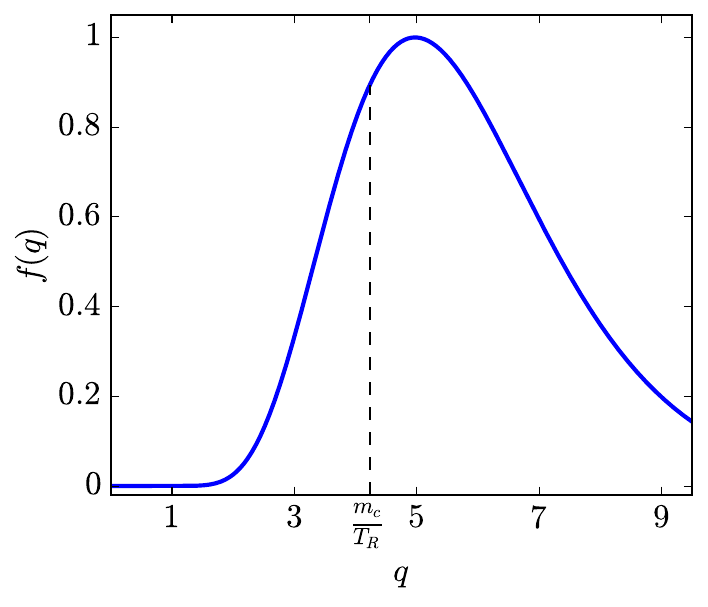}
    \caption{Comoving momentum distribution for $T_R = 300\,\mathrm{MeV}$, including only the charm quark contribution for an instantaneous reheating temperature profile. The distribution is normalized to unity at its maximum.}
    \label{fig:psd_instant}
\end{figure}

For the instant-reheating profile the integral \eqref{eq:psd} is saturated at $T \simeq T_R$, and Eqs.~\eqref{eq:Asymp_1}--\eqref{eq:Asymp_2} give
\begin{align}
    &q \ll \frac{m_f^2}{T_R^2} \, : \qquad 
    f_\chi(q) \propto q^{3/2} \exp\left(-\frac{m_f^2}{qT_R^2}\right)\,, \\[4pt]
    &q \gg \frac{m_f^2}{T_R^2} \, : \qquad 
    f_\chi(q) \propto q\, e^{-q}\,.
\end{align}
The result is shown in Fig.~\ref{fig:psd_instant}. The distribution is strongly non-thermal: the low-momentum modes are exponentially suppressed and the spectrum peaks at $q \sim m_f/T_R \gg 1$. The suppression has a simple origin. DM is produced with a physical momentum set by the fermion mass, $p \sim m_f$, which then redshifts as $T/T_R$. Since production is cut off at $T_R$, the particles do not have enough time to redshift down to $q \sim 1$, and the infrared modes are depleted. The resulting large comoving momentum, $q \sim m_f/T_R$, is what makes the relic warm and drives the Lyman-$\alpha$ bound.

% =========================================================
\subsection{Flat temperature profile}
\label{subsec:PSD_Flat_T}

For a constant temperature $T = T_R$ before reheating, the distribution receives an additional contribution from the pre-reheating epoch,
\begin{equation}
    f(q)^\mathrm{Flat-T} = f(q) + \Delta f(q)\,,
\end{equation}
where $f(q)$ is the post-reheating part discussed above and \cite{Feiteira:2026qme}
\begin{equation}
    \Delta f(p,a_R) = \frac{1}{p a_R} \int_0^{a_R} \frac{da'}{H(a')} \mathcal{C}(p',T_R)\,,
    \qquad 
    p' = p \frac{a_R}{a'}\,,
\end{equation}
with $T = T_R$ and $H(a) \propto a^{-2}$ throughout. The integral is controlled by a narrow region around the saddle point $a'_* \sim a_R\, p/m_f$, the time at which particles of momentum $p$ are most efficiently produced.

\begin{figure}
    \centering
    \includegraphics[width=0.5\linewidth]{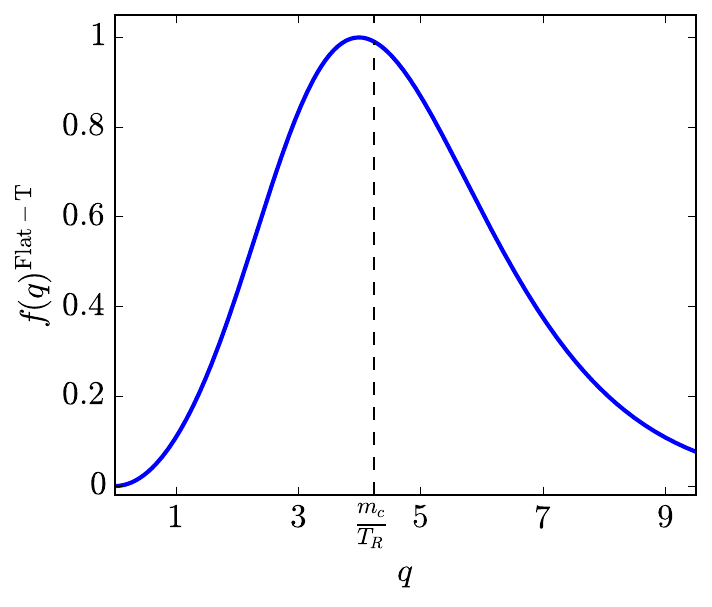}
    \caption{Comoving momentum distribution for $T_R = 300\,\mathrm{MeV}$, including only the charm quark contribution, assuming a flat temperature profile before reheating. The distribution is normalized to unity at its maximum.}
    \label{fig:psd_flat}
\end{figure}

At small momenta, $q \ll 1$, this yields $\Delta f(q) \propto q^2$, up to an overall factor $\propto e^{-2m_f/T_R}$, so that
\begin{equation}
    f(q)^\mathrm{Flat-T} \propto q^2 \qquad \mathrm{for} \qquad q \ll 1\,,
\end{equation}
in sharp contrast with the exponential suppression of the instant-reheating case. At large momenta, $q \gg 1$, late-time production takes over and the same fall-off $\Delta f(q) \propto e^{-q}$ is recovered. The constant-temperature phase thus fills in the infrared, trading the exponential cutoff for a power law, as shown in Fig.~\ref{fig:psd_flat}. The low- and high-momentum tails are now set by the pre- and post-reheating epochs, respectively. In both profiles the two tails scale with different powers, so the full distribution is not captured by the standard $\alpha\beta\gamma$-parametrization $f(q) \propto q^\alpha e^{\beta q^\gamma}$ \cite{Bae:2017dpt}.

%%%%%%%%%%%%%%%%%%%%%%%%%%%%
\section{Parameter space analysis}
%%%%%%%%%%%%%%%%%%%%%%%%%%%%

\subsection{Instant reheating temperature profile}

\begin{figure}
    \centering
    \includegraphics[width=0.5\linewidth]{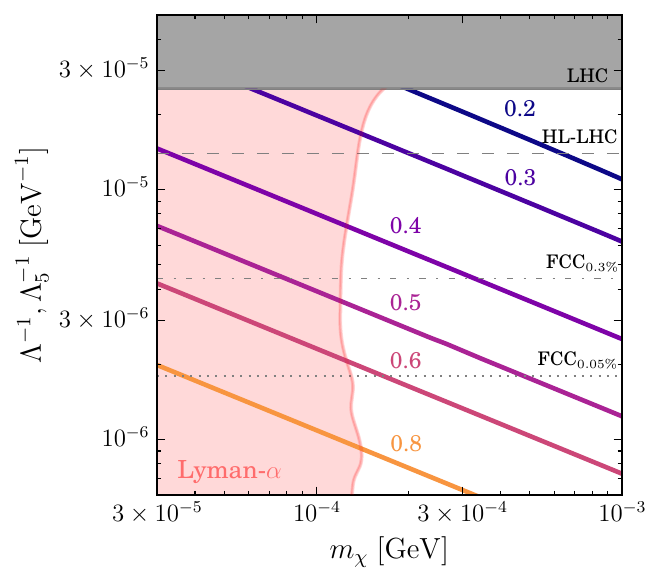}
    \caption{Parameter space of the Higgs portal fermionic DM at low masses, for an instant reheating temperature profile. Along the colored lines the observed DM abundance is reproduced, for a given $T_R$ in GeV. The shaded areas are excluded by the LHC and Lyman-$\alpha$ constraints, while the dashed line shows the projected sensitivity of HL-LHC. Sensitivities of the FCCs are shown by the dashed-dotted and dotted lines.}
    \label{fig:parameter_instant}
\end{figure}

\autoref{fig:parameter_instant} shows our parameter space analysis. We restrict it to $m_\chi \lesssim 1\,\mathrm{MeV}$, since for larger masses the fermionic DM is cold\footnote{The region $m_\chi \geq 1$ MeV has already been studied in Ref.~\cite{Lebedev:2024mbj}.}. Along the colored curves the observed relic abundance is reproduced for the indicated reheating temperature $T_R$ in GeV. The grey region is excluded by the LHC search for invisible Higgs decay ($\mathrm{BR}_\mathrm{inv} \lesssim 10\%$ \cite{ATLAS:2023tkt}), and the dashed lines mark the projected reaches of HL-LHC and FCC. For HL-LHC we take $\mathrm{BR}_\mathrm{inv} \lesssim 3\%$, while for FCC we use two benchmarks: a conservative one, $\mathrm{BR}_\mathrm{inv} \lesssim 0.3\%$ \cite{FCC}, used in earlier studies, and an updated one, $\mathrm{BR}_\mathrm{inv} \lesssim 0.05\%$ \cite{FCC2}.

The pink region is excluded by the Lyman-$\alpha$ bound. To obtain it, we map the Lyman-$\alpha$ constraints of warm DM (WDM) to our non-thermal DM using the area criterion, detailed in Refs.~\cite{DEramo:2020gpr,Decant:2021mhj}. We apply the thermal warm DM mass bound of $5.3\,\mathrm{keV}$ \cite{Irsic:2017ixq}. The linear matter power spectra, for both the thermal WDM and our non-thermal DM, are computed with the CLASS tool \cite{Murgia:2017lwo,Blas:2011rf,Lesgourgues:2011rh}, following the implementation described in \cite{Feiteira:2026qme}. As a result, masses below $120 - 177 \,\rm keV$ are excluded. The weakest bound, $m_\chi >120 \,\rm keV$, occurs at $T_R = 0.5 \,\rm GeV$, and the strongest, $m_\chi >177 \, \rm keV$, at $T_R \approx 0.2 \, \rm GeV$.

Compared to the scalar Higgs portal \cite{Feiteira:2026qme}, the reheating temperatures that give the correct abundance in the allowed region are much higher, well above the QCD phase transition. The reason is the $s$-dependence of the fermionic cross section, which suppresses production near the kinematic threshold (low relative velocity of the initial fermions). At low $T_R$ this suppression is stronger than in the scalar case, so a larger coupling $\Lambda^{-1}$ is needed to match the relic abundance at a given $T_R$. The LHC bound on $\Lambda^{-1}$ then pushes the allowed region to $T_R \gtrsim 0.2\,\mathrm{GeV}$, above the confinement scale.

\begin{figure}[htbp]
    \centering
    \includegraphics[width=\textwidth]{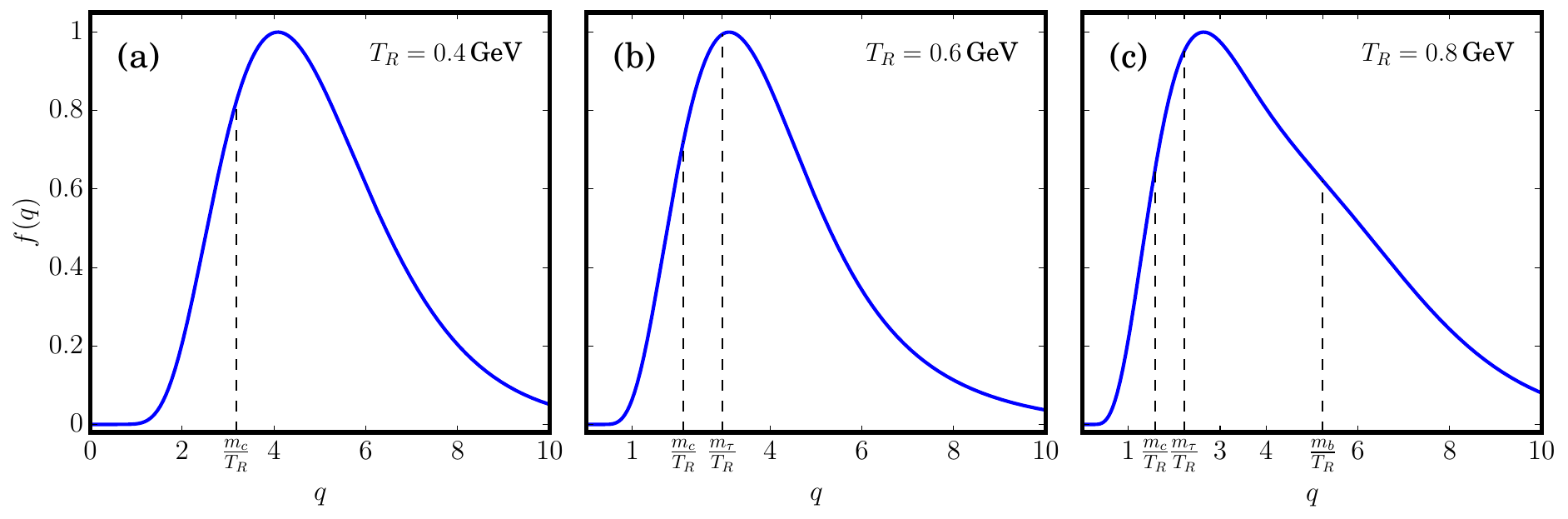}
    \caption{Comoving momentum distribution of DM including all production channels, for different reheating temperatures $T_R$ and assuming instant-like reheating. The dotted lines mark $m_f/T_R$ for the relevant channels. For convenience, $f(q)$ is normalized to unity at the maximum.}
    \label{fig:distributions}
\end{figure}

The most striking feature of \autoref{fig:parameter_instant} is the non-monotonic Lyman-$\alpha$ bound. To understand it, we show in \autoref{fig:distributions} the comoving momentum distribution of DM for different reheating temperatures, and follow which channel dominates production as $T_R$ changes. The point is that DM is produced with a momentum set by the mass of the annihilating fermion, so its comoving momentum peaks near $q \sim m_f/T_R$ (see the dotted lines in \autoref{fig:distributions}). A heavier dominant channel therefore gives a warmer spectrum and a tighter free-streaming bound.

For $0.2\,\mathrm{GeV} \lesssim T_R \lesssim 0.5\,\mathrm{GeV}$ the charm channel dominates and the distribution looks like \autoref{fig:distributions}a. Here the bound tightens as $T_R$ drops, because a lower $T_R$ both raises the peak momentum $m_c/T_R$ and leaves less time for the spectrum to redshift before structure formation.

Around $T_R \approx 0.5\,\mathrm{GeV}$ the trend reverses: the bound now tightens as $T_R$ grows. The charm channel still dominates, but the tau channel becomes important and, being heavier, injects DM at higher momentum. The peak of $f(q)$ moves up and the hard tail grows (see \autoref{fig:distributions}b for $T_R = 0.6\,\mathrm{GeV}$), so only heavier masses survive. By $T_R \approx 0.7\,\mathrm{GeV}$ the bound relaxes again, since the rise in $T_R$ softens the spectrum more than the tau channel hardens it.

Between $T_R \approx 0.7\,\mathrm{GeV}$ and $T_R \approx 0.8\,\mathrm{GeV}$ the bound tightens once more. The bottom quark, much heavier than charm and tau, now contributes and fills the high-momentum tail (\autoref{fig:distributions}c, $T_R = 0.8\,\mathrm{GeV}$). Above $T_R \approx 0.85\,\mathrm{GeV}$ no new channels open, and the bound weakens steadily up to $T_R \approx 1\,\mathrm{GeV}$ as the spectrum softens.

This non-monotonic Lyman-$\alpha$ bound is another clear difference from the scalar case. There, the same two channels, pion and muon, dominate the whole allowed region, so the bound simply weakens with $T_R$. For fermionic DM, the quark and lepton channels switch on at well-separated values of $T_R$, each one briefly hardening the spectrum and producing the oscillations seen in \autoref{fig:parameter_instant}.

\subsection{Flat temperature profile}

\begin{figure}
    \centering
    \includegraphics[width=0.5\linewidth]{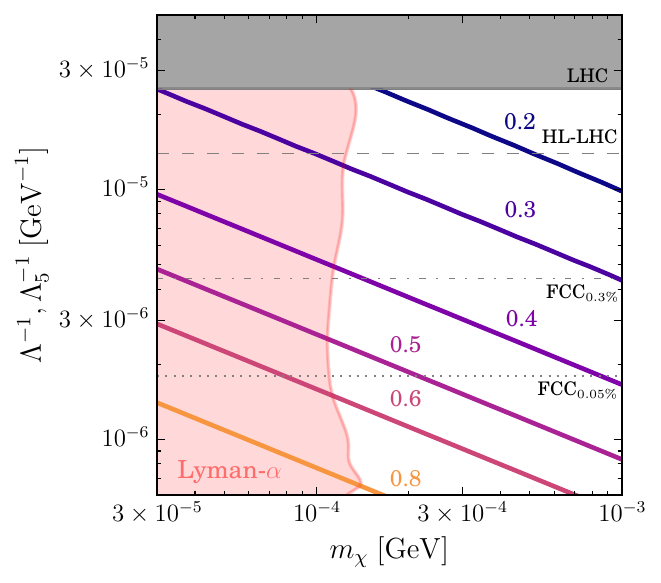}
    \caption{Parameter space of the Higgs portal fermionic DM at low masses, for a flat temperature profile before reheating. Along the colored lines the observed DM abundance is reproduced, for a given $T_R$ in GeV. The shaded areas are excluded by the LHC and Lyman-$\alpha$ constraints, while the dashed line shows the projected sensitivity of HL-LHC. Sensitivities of the FCCs are shown by the dashed-dotted and dotted lines.}
    \label{fig:parameter_flatT}
\end{figure}

The results for the flat temperature profile are shown in \autoref{fig:parameter_flatT}, with the same conventions as in \autoref{fig:parameter_instant}. The constraints are weaker than in the instant reheating case. The reason is the infrared shape of the distribution: as shown in Section~\ref{subsec:PSD_Flat_T}, the constant-temperature phase replaces the exponential low-momentum suppression by a power law, $f(q) \propto q^2$. More DM is then produced at low momentum, which lowers the average free-streaming length and allows lighter DM for a given $T_R$. Masses below $95 - 140 \,\rm keV$ are excluded. The weakest bound, $m_\chi >95 \,\rm keV$, occurs at $T_R \approx 0.2 \,\rm GeV$, and the strongest, $m_\chi > 140 \, \rm keV$, at $T_R = 0.8 \, \rm GeV$.

The Lyman-$\alpha$ bound again oscillates, for the same reason as before: the successive activation of heavier channels. One feature is specific to this profile. Between $T_R \approx 0.2\,\mathrm{GeV}$ and $T_R \approx 0.25\,\mathrm{GeV}$ the bound tightens as $T_R$ grows, contrary to the naive expectation. This happens because the muon channel matters at these low reheating temperatures and fades as $T_R$ increases. The muon, being much lighter than the charm, produces low-momentum DM. This effect was absent in the instant reheating profile, where the exponential suppression at low momentum erases any contribution to the soft part of the spectrum.

\section{Conclusion}

We have studied the Higgs portal to fermionic dark matter through both CP-even and CP-odd interactions, $\mathcal{L}^{\chi}_\mathrm{eff} = \frac{m_f}{\Lambda m_h^2}\,\bar f f\,\bar\chi\chi$ and $\mathcal{L}^{\chi\gamma_5}_\mathrm{eff} = \frac{m_f}{\Lambda_5 m_h^2}\,\bar f f\,\bar\chi i\gamma_5\chi$, in the context of freeze-in at stronger coupling, for DM masses below $1\,\mathrm{MeV}$ and reheating temperatures below $800\,\mathrm{MeV}$. In this regime, the part of the parameter space allowed by LHC requires $T_R \gtrsim 0.2 \,\rm GeV$. Since this is above the QCD phase transition, the main interactions producing DM are the charm, tau, and bottom channels. The production rate is Boltzmann suppressed for $T_R < m_c,\, m_\tau, \,m_b$, so a larger Higgs-DM coupling is needed to match the relic abundance. As a result, this freeze-in DM can be probed at colliders, unlike standard high-temperature freeze-in.

The correct relic abundance is reproduced for an effective scale $\Lambda$ between $3\times10^5$ and $10^6 \,\rm GeV$. Part of this range can be tested through invisible Higgs decay at HL-LHC and FCC. Freeze-in at stronger coupling thus brings fermionic non-thermal dark matter within reach of colliders.

The Lyman-$\alpha$ bound strongly constrains the fermionic DM mass, excluding masses below about $150 \,\rm keV$. This is tighter than in standard high-temperature freeze-in, because DM from low-temperature freeze-in is strongly non-thermal, produced with a comoving momentum of order $m_c/T_R \gg 1$.  It is also tighter than for scalar dark matter from the same mechanism, because the dominant channels involve heavier particles. For scalar DM the dominant channels over most of the parameter space are pion and muon scattering, while for fermionic DM it is charm scattering. The typical momentum is therefore $m_c \gg m_\pi, \, m_\mu$, giving a harder spectrum, which requires a lower mass to still allow small-scale structure to form.  In general, we also find that the resulting momentum distribution in a low reheating-temperature scenario is not captured by the usual $\alpha\beta\gamma$ parametrization.

The Lyman-$\alpha$ curve for fermionic dark matter shows small oscillations, absent for scalar dark matter. These come from different production channels becoming relevant at different reheating temperatures.

\paragraph{Acknowledgements}
We are grateful to Oleg Lebedev for the important discussions.
V.O. is directly funded by FCT through the doctoral program grant with the reference PRT/BD/154629/2022 (\url{https://doi.org/10.54499/PRT/BD/154629/2022}).

\bibliographystyle{JHEPfixed}
\bibliography{references}

\end{document}